# A Prototype System for Controlling a Computer by Head Movements and Voice Commands

Anis Ismail, Abd El Salam AL HAJJAR and Mohammad HAJJAR

Lebanaise University, Saida, BP 813, LIBAN

anismaiil@ul.edu.lb
abdsalamhajjar@hotmail.com
M_hajjar@ul.edu.lb

**ABSTRACT**

*This paper introduces a new prototype system for controlling a PC by head movements and also with voice commands. Our system is a multimodal interface concerned with controlling the computer. The selected modes of interaction are speech and gestures. We are seeing the revolutionary of computers and information technologies into daily practice. Healthy people use keyboard, mouse, trackball, or touchpad for controlling the PC. However these peripheries are usually not suitable for handicapped people. They may have problems using these standard peripheries, for example when they suffer from myopathy, or cannot move their hands after an injury. Our system has been developed to provide computer access for people with severe disabilities. This system tracks the computer user's Head movements with a video camera and translates them into the movements of the mouse pointer on the screen and the voice as button presses. Therefore we are coming with a proposal system that can be used with handicapped people to control the PC.*

## 1. INTRODUCTION

With the development of information technology in our society, we can expect that computer systems to a larger extent will be embedded into our environment. These environments will impose needs for new types of human-computer interaction, with interfaces that are natural and easy to use. In particular, the ability to interact with computerized equipment without need for special external equipment is attractive.

Today, the keyboard, the mouse and the remote control are used as the main interfaces for transferring information and commands to computerized equipment. In some applications involving three-dimensional information, such as visualization, computer games and control of robots, other interfaces based on trackballs, joysticks and datagloves are being used. In our daily life, however, we humans use our vision and hearing as main sources of information about our environment. Therefore, one may ask to what extent it would be possible to develop computerized equipment able to communicate with humans in a similar way, by understanding visual and auditive input.

For many people with physical disabilities, computers form an essential tool for communication, environmental control, education and entertainment. However, access to the computer may be made more difficult by a person's disability. A number of users employ head-operated mice or joysticks in order to interact with a computer and to type with the aid of an on-screen keyboard. Head-operated mice can be expensive. In the UK, devices that require the users to wear no equipment on their heads, other than an infrared reflective dot, for example Orin Instrument's HeadMouse [22] and Prentke Romich's HeadMaster Plus [25]. Other devices are cheaper, notably Granada Learning's Head Mouse [7] [9] [10], Penny and Gilles' device, Mouse Enhancer for paraplegics [17], EyeGaze System [11] and No Hands Mouse [19]. However, these systems require the user to wear a relatively complex piece of equipment on their head, an infrared transmitter and a set of mercury tilt switches respectively.

People with severe disabilities who retained the ability to rotate their heads have other assistive technology options. For example, there are various commercial mouse alternatives. Some systems use infrared emitters that are attached to the user's glasses, head band, or cap. Other systems place the transmitter over the monitor and use an infrared reflector that is attached to the user's forehead or glasses. The user's head movements control the mouse cursor on the screen.

As we examine vision-based cursor control systems available on the market [3] [28] [16] [26] [20] or research literature [2] [4] [5] [6] [8] [12] [27] [24] [26], we note that in order to provide a user with the knowledge on how the face is detected by the camera, these interfaces use a separate window somewhere on a screen in addition to normal cursor, which shows the capture video image of face with the results on vision detection overlaid on top of it. This is how the visual feedback is provided for Camera Mouse [3], QualiEye [26], IBM Head Tracker, which are the examples of the commercially sold vision based computer control programs that have been tested by several handicapped people in many Health Center. The drawback of this visual feedback is that the user has to look both at the cursor (to know where/how to move it, e.g. to open a Windows menu) and at the image showing the results captured by the video camera (to know how to move his head in order to achieve the desired cursor motion). Since a user cannot view two different locations at the same time, this creates a problem for the user. Furthermore, having an additional window produces another problem. It occludes other window applications, making the windows desktop more cluttered and less organized. Gorodnichy [13] resolved this problem by introducing a new concept, called Perceptual Cursor, which serves both the purpose of marking a position (as normal cursor) and the purpose of providing a user with the feedback on how remote user motions are perceived by a sensor. As such, Perceptual Cursor does not replace the regular cursor, but rather is used in the interface in addition to it, taking its functionality only when requested by the user. For voice-based computer control system available in the market, we can find Control Your PC with Your Voice [29], Nuance Voice Control 2.0 [21] which are an easy software solution to enable you to control your computer, dictate emails and letters using a minimum of keystrokes or mouse clicks.

Delphian Desktop [32] predicts the destination of the cursor based on initial movement and rapidly 'flies' the cursor towards its target. Although these techniques were designed for single monitor conditions, they can be easily tailored for multi-monitor setups. Head and eye tracking techniques were proposed to position the cursor on the monitor of interest [33, 34]. This reduces significant mouse trips but at the cost of access to expensive tracking equipment. Benko et al. propose to manually issue a command (i.e. a button click) to ship the mouse pointer to a desired screen [35]. Ninja cursor [36] proposes a technique to improve the performance of target acquisition, particularly on large screens. This technique uses multiple distributed cursors to reduce the average distance to targets. Each cursor moves synchronously following mouse movement.

Villar et. al [37] explore the possibilities for augmenting the standard computer mouse with multi-touch capabilities so that it can sense the position of the user's fingers and thereby complement traditional pointer-based desktop interactions with touch and gestures. They present five different multi-touch mouse implementations, each of which explores a different touch sensing strategy, which leads to differing form-factors and hence interaction possibilities. In addition to the detailed description of hardware and software implementations of their prototypes, they discuss the relative strengths, limitations and affordances of these different input devices as informed by the results of a preliminary user study.

LensMouse [38] presents a novel device that embeds a touch-screen display – or tangible 'lens' – onto a mouse. Users interact with the display of the mouse using direct touch, whilst also performing regular cursor-based mouse interactions. Yang et al. demonstrate some of the unique capabilities of such a device, in particular for interacting with auxiliary windows, such as toolbars, palettes, pop-ups and dialog-boxes. By migrating these windows onto LensMouse, challenges such as screen real-estate use and window management can be alleviated. In a controlled experiment, they evaluate the effectiveness of LensMouse in reducing cursor movements for interacting with auxiliary windows. they also consider the concerns involving the view separation that results from introducing such a display-based device.

The goal of this paper is to present a new system for non-contact interface that is both low cost and also does not require the user to wear any equipment. The most obvious way of doing this is to use a camera, interfaced to a PC. Our system will track the movement of the user; it will translate head movements into cursor movements, and the voice as button presses. There have been many attempts at creating cursor control systems utilizing head movements. In our case the speech recognition system must also provide a mechanism for replacing the mouse buttons. This system targets people suffering from reduced mobility as well as specific professionals operating in constraining situations. This design allows using simple head movements to perform basic computer mouse operations, such as moving the mouse cursor on a computer screen. To improve functionality, the system uses voice recognition to execute repetitive movements, such as clicks and double-clicks. The strength of this system is the combination of the vision controlling the cursor with the voice commands. So our system is a multimodal interface concerned with controlling the computer.

Multimodal interaction provides the user with multiple modes of interfacing with a system beyond the traditional keyboard and mouse input/output. The most common such interface combines a visual modality (e.g. a display, keyboard, and mouse) with a voice modality (speech recognition for input, speech synthesis and recorded audio for output).

The advantage of multiple modalities is increased usability: the weaknesses of one modality are offset by the strengths of another. On a mobile device with a small visual interface and keypad, a word may be quite difficult to type but very easy to say. Consider how you would access and search through digital media catalogs from these same devices or set top boxes. And in one real-world example, patient information in an operating room environment is accessed verbally by members of the surgical team to maintain an antiseptic (sterile) environment, and presented in near real-time aurally and visually to maximize comprehension.

Multimodal user interfaces have implications for accessibility [31]. A well-designed multimodal application can be used by people with a wide variety of impairments (disabilities). Patients with advanced muscular dystrophy and multiple sclerosis revealed that their primary interest was assistive software that would enable them to have access to communication and information, i.e., text entry, email, and web browsing [1]. Visually impaired users rely on the voice modality with some keypad input. Hearing-impaired users rely on the visual modality with some speech input. Other users will be "situationally impaired" (e.g. wearing gloves in a very noisy environment, driving, or needing to enter a credit card number in a public place) and will simply use the appropriate modalities as desired. On the other hand, a multimodal application that requires users to be able to operate all modalities is very poorly designed.

The remainder of this paper is organized as follows. In section 2, we represent the problematic. Our system architecture is described in the section 3. In the last section, we conclude the main results of this paper

## 2. PROBLEMATIC

With advances made in computer vision, a lot of research effort has been focused recently on designing a user Interfaces based on computer vision, which are the systems that use a videocamera to detect the visual cues of the user, such as the motion of the face, to control the cursor of a computer. One of the next major steps in the advancement of computing devices is not only making them faster, but making them more interactive, responsive, and accessible to the user. The achievement of the natural human-computer interaction requires the use of the modalities that we ourselves use to communicate. Some user interfaces combine natural human capabilities of communication, motor, cognitive, and perceptual skills with computer I/O devices, machine perception, and reasoning.

One of the most important tasks for the system user interface is seen in enabling a hands-free control of the cursor. The reason for this is that the cursor, which is normally controlled by hand-operated input devices such as mouse, joystick, track pad and track ball, serves as a prime link between the user and the computer commands the user wants to execute. In particular, the cursor is used to choose the items in Windows menu, position oneself in the desired location in an editor or other window program, highlight a text and objects on a screen, switch focus of attention between the windows, activate and deactivate programs, etc.

Moving the cursor is also the least constrained and required the least skill technique the user can perform when interacting with a computer. Other, more sophisticated ways of interacting with a computer, such as typing with a keyboard, can be simulated using the cursor motion and the cursor-operated on-screen tools, such as on-screen keyboard.

This explains why for users with hand motion deficiency, the ability to move the cursor with the motion of the face appears to be a long-waited solution for their computer needs, while for all other users hands-free face-motion-controlled cursor is welcomed as an additional degree of computer control much needed when hands are busy with other tasks [15].

Regardless of the significant advances recently made in vision-based face detection, of which we acknowledge the following techniques as the ones contributing the most to the field: developed for automatic detection of faces [24] and eyes [14], 3D-face-model-based tracking [30] and robust sub-pixel precision nose tracking [12] should be specially acknowledged, the desired hands-free face-tracking-based cursor control has not been achieved yet.

The problem is that, while making it possible to receive user commands remotely, the user interface, based on computer vision, introduces one major problem — the absence of "touch" with the cursor. By holding a mouse, a user not only controls the program, but also keeps the knowledge of where the mouse is. No matter how robust the user interface is, it can lose the user; it might be even more appropriate to say that a user loses the interface. What a user needs in order to control a cursor is not a more precise or robust motion detection technique, but we need a reference for the coordinates of the cursor on a screen when the user lose the cursor of the mouse.

The underlying problematic to the control of the computer environment by persons with motor deficiencies is this: how can persons who are physically incapable of manipulating the keyboard or the mouse of the computer, access the different services and applications distributed by the software publishers? The solution that we have developed a system in a coherent way to control computer with different accessibility modalities such as head movement ant speech recognition.

## 3. SYSTEM OVERVIEW

Our system is a mouse emulator system based on the facial movement of the user. This software program developed is a small program (around 1.1 MB) which interprets a specific set of voice commands from the user. The head movement sensing system is augmented with a voice recognition program installed on the user's PC. The user's head movements will control a mouse cursor on the PC screen, just like a standard mouse pointing device. The user's voice commands allow performing mouse clicks and other similar operations.

A Webcam is placed in front of the user, focusing on the user's face. A motion extraction algorithm based on RGB, which is user independent, is used to extract the facial motion from the video. This motion is used to move the mouse pointer that is controlled in a fashion relatively similar to standard mouse devices. This system can be used with great accuracy even when the user has exiguous cephalic motion control.

The voice recognition program installed on the computer interprets the user's voice commands and translates them into commands such as "click" (or "left click"), "right click" or "double click". A microphone system transmits the voice commands from the user to the computer sound input.

### 3.1 Algorithm

#### 3.1.1 Head detection

To recognize the target head gesture acknowledgements, we first must locate the face of the user in the image. Necessary requirements of the face detection algorithm are that it be real-time and robust to various lighting conditions, background noise, and skin color. The approach used in this research first locates the face by detecting skin-face areas. So, we have chosen to use color information to identify the face. Our aim was to use the simplest reliable algorithm that will satisfy our requirements. We do not use other methods of face detection as we supposed them to be overly complex for this application, and although they might yield correct results at an acceptable frame rate, we do not believe that they will leave sufficient processing resources to allow any useful software to be used without an unacceptable degradation in performance.

Our system precise the face that contains skin by extracting the red, green and blue (RGB) components of each pixel. The RGB values were converted to normalized red, green and blue.

R = c Mod 256

G = (c / 256) Mod 256

B = (c / 256 / 256) Mod 256

c = Picture-Point(X, Y): represents the image pixels.

We have demonstrated that it is possible to identify a range of colors (RGB) that include all skin colorations. Of course, these colors will not be unique to skin samples, the colors of other objects will also be found in this range. We have found that certain colors of paint like the background. These pixels are readily removed using background subtraction. So the background should be blank.

The output of this stage of the processing cycle is a rectangle around the face that indicates the pixels that are of a color consistent with the skin of the face.

### 3.1.2. Speech Recognition

Speech recognition requires the computer to accept spoken words as input and interpret what has been spoken. To make the job of understanding speech easier for the computer, a method of speech input called command and control is used.

Speech Recognition is technology that allows a computer to identify the words that a person speaks into a microphone. We used Microsoft Agent version 2.0 that provides a library for more natural ways for people to communicate with their computers. And also we used The Lernout & Hauspie TruVoice Text-to-Speech (TTS) [18] Engine that provides speech output capabilities for Microsoft Agent so we can hear what the characters are saying through your sound speakers.

The commands available to the user are the following: "left click" (or just "click"), "double left click" (or just "double click"), "right click", "double right click". The user can also keep a button pressed so as to highlight a group of objects. The command "down", "up" change the selection area of the mouse per example in the menu (File, Edit, Insert…) and we can select also the menu with the voice command. This set of available commands allows executing meaningful tasks on the computer since all the main mouse click operations are available.

Our software program uses generic speech recognition. This means that the computer doesn't have to be "trained" to understand a specific user's voice. This allows greater flexibility, since the system is easily accessible to new users.

### 3.2. System settings

### 3.2.1. Face detection settings

Our system requires an initialization procedure. The Camera's position is adjusted so as to have the user's face in the center of the camera's view. Then, the feature to be tracked, in this case the face is selected with the value of the RGB in the title of this window (See Figure 1).

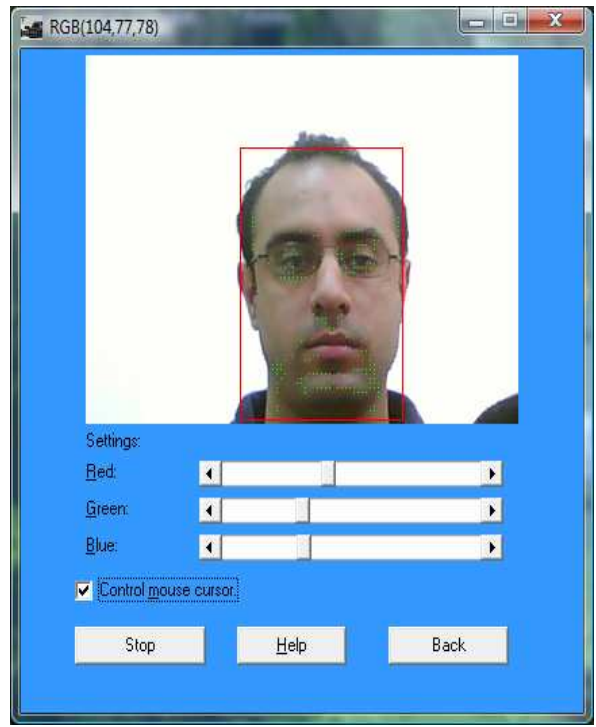

Figure 1  Face detection initialization

### 3.2.2. Voice commands setting

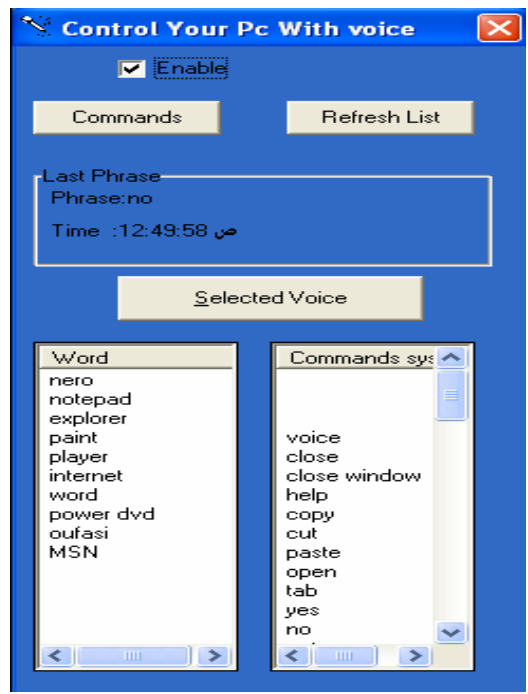

Figure 2  Voice command initialization

In figure3, the users see a list of services that they can control and access them through voice commands. A list that is maintained by this interface is to show the user what applications are available for use (See Figure 2). Applications that are currently in use are in the left frame and commands system in the right frame. We can add or remove applications to this list some applications and also voice commands.

### 3.3. Demo

In this demo, we lunch the internet with voice commands. After checking the enable to active the voice commands, per example to lunch internet explorer application user uses saying "internet", immediately a confirmation dialog box appear on which the saying command will be written (See Figure 3). After saying "yes", Internet Explorer appears then we can control it by voice commands by saying "file, edit…" and by "up, down, ok (for enter) …" the options in the menu (file, edit, view…) (See Figure 4).

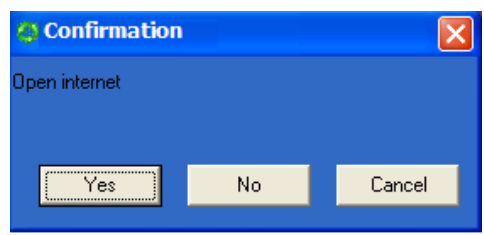

Figure 3  Confirmation

### 3.4. Performance

We evaluated the performance of our system in several different situations and lighting conditions. We also tried it on different people and got good results.

First we tried our system to see whether the face detection was continuously being tracked and whether the template updating procedure that we suggested gives good results and does not drift away with time. Our tests were aimed at evaluating the precision, speed and accuracy of the face based tracking system. Our tests showed that processing speed, problems in many previous approaches, were not a problem in our system. Unfortunately, our system does not work well in the presence of dark background that should be blank.

## 4. CONCLUSION

The multimodal system is aimed for the disabled people, which need other kinds of interfaces than ordinary people. In the developed system the interaction between a user and a computer is performed by voice and head movements. To process these data streams the modules of speech recognition and head tracking were developed. This system was applied for hands-free operations with Graphical User Interface in such tasks as Internet communications and lunching applications. We showed theoretically and practically that this technology could be used to operate computers hands-free. Our prototype exhibits accuracy and speed, which are sufficient for many real time applications and which allow handicapped users to enjoy many computer activities. The experiments have shown that in spite of some decreasing of operation speed the multimodal system allows working with computer without using standard mouse and keyboard. Thus the developed assistive multimodal system can be successfully used for hands-free PC control for users with disabilities of their hands or arms.

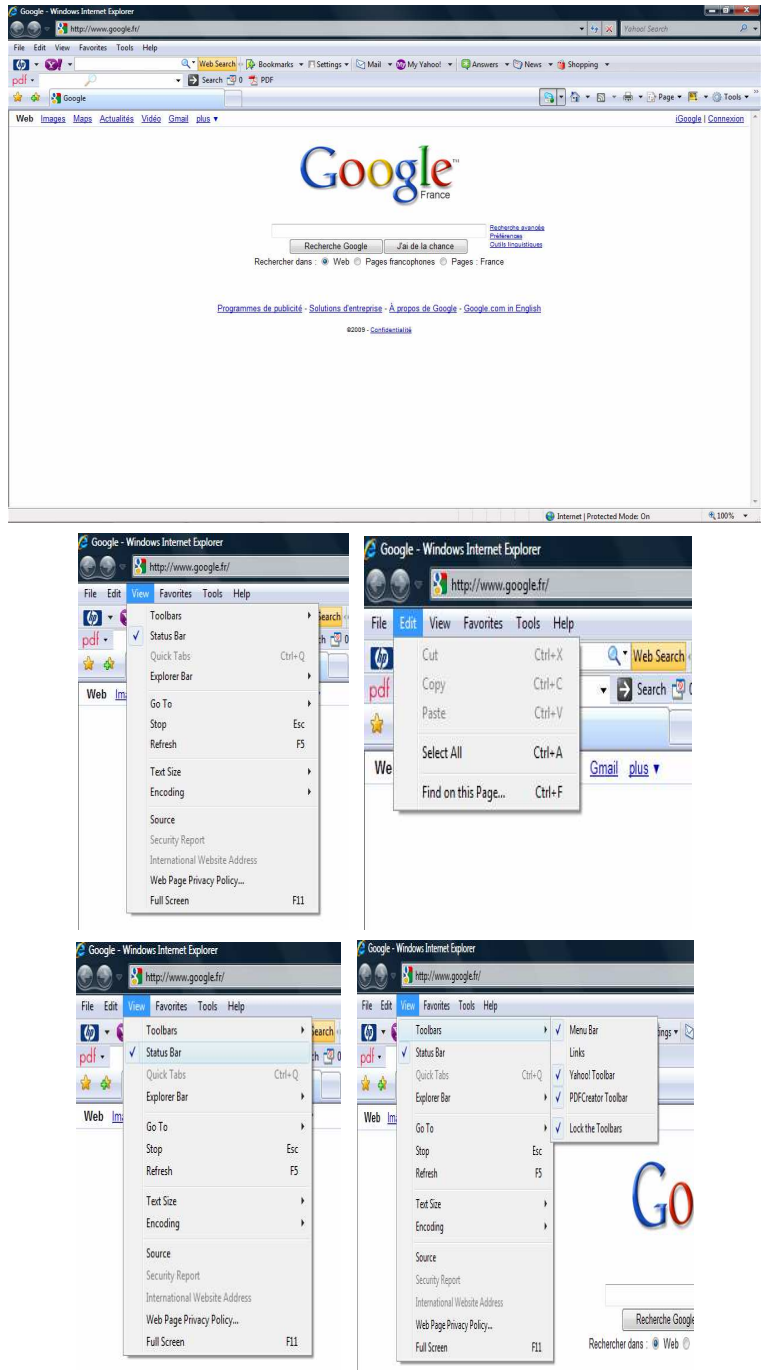

Figure 4 Controlling internet explorer with voice commands